\documentclass[twocolumn,prl,superscriptaddress,showpacs,citeautoscript,floatfix]{revtex4-1}

\usepackage{graphicx}
\usepackage{dcolumn}
\usepackage{bm}
\usepackage{flafter}
\usepackage{lettrine}
\usepackage{amsmath}
\usepackage{amssymb}
\usepackage{color}
\usepackage{epsfig}
\usepackage{float}
\usepackage{type1cm}
\usepackage{tabu}
\graphicspath{{eps+pdf/}}

\begin{document}

\title{High-pressure structural study of $\alpha$-Mn: solving a three decades-old mystery.}
\author{Logan K. Magad-Weiss}
\affiliation {Lawrence Livermore National Laboratory, Physical and Life Sciences Directorate,
Livermore, California 94550, USA}
\author {Adebayo A. Adeleke}
\affiliation{Department of Physics and Engineering Physics, University of Saskatchewan, Saskatoon Saskatchewan, S7N 5E2, Canada}
\author{Eran Greenberg}
\affiliation{Center for Advanced Radiation Sources, University of Chicago, Chicago, IL 60637, USA}
\author{Vitali B. Prakapenka}
\affiliation{Center for Advanced Radiation Sources, University of Chicago, Chicago, IL 60637, USA}
\author{Yansun Yao}
\email{yansun.yao@usask.ca}
\affiliation {Department of Physics and Engineering Physics, University of Saskatchewan, Saskatoon Saskatchewan, S7N 5E2, Canada}
\date{\today}
\author{Elissaios Stavrou }
\email{stavrou1@llnl.gov}
\affiliation {Lawrence Livermore National Laboratory, Physical and Life Sciences Directorate,
Livermore, California 94550, USA}

\begin{abstract}
Manganese, in the $\alpha$-Mn structure,  has been studied using synchrotron powder x-ray diffraction  in a diamond anvil cell up to 220 GPa at room temperature combined with density functional calculations (DFT).  The experiment reveals an extended pressure stability of the $\alpha$-Mn phase up to the highest pressure of this study, in contrast with previous experimental and theoretical studies. On the other hand, calculations reveal  that the previously predicted  hcp-Mn phase becomes lower in enthalpy than the $\alpha$-Mn phase above 160 GPa. The apparent discrepancy is explained due to a substantial electron transfer between Mn ions, which stabilizes the $\alpha$-Mn phase through the formation of ionic bonding between monatomic ions under pressure.
\end{abstract}
\maketitle

\section{Introduction}
At ambient conditions manganese crystalizes in the, unparalleled among the elements, complex $\alpha$-Mn structure (S.G. $I-43m$ (217),{} Z=58). It was proposed \cite{Brewer1968} that the complexity (with a very large number of atoms in the cubic unit cell)  of this structure originates from a combination of the presence of three electronic levels  with comparable stability and magnetic ordering, resulting in different  atomic sizes. Thus, Mn atoms with different atomic sizes are accommodated in different  crystallographical sites, $i.e.$ Mn atoms behave
as if they are different atoms with different sizes \cite{Brewer1968,Fujihisa1995}. Indeed, the $\alpha$-Mn phase is analogous to the $\chi$-phase  in binary or ternary intermetallic systems \cite{Fujihisa1995}.

One would expect that under pressure the suppression of the magnetic moments \cite{Johansson1998} will result in a pressure induced phase transition towards a much simpler crystal structure with equal atomic size. Previous experimental x-ray diffraction (XRD) studies \cite{Fujihisa1995,Takemura1988} reported a, somewhat surprising, stability of the $\alpha$-Mn phase up to at least 150 GPa. At higher pressure, a  phase transition was reported by Fujihisa and Takemura \cite{Takemura1988} based on the appearance of a new low intensity Bragg peak above 165 GPa, that coexisted with the main Bragg peak of the $\alpha$-Mn phase up to 190 GPa. After considering three simple metallic structures (bcc, fcc, and hcp), a bcc structure was suggested as the high-pressure phase based on the lower volume collapse during the transition of the $\alpha$-Mn to the bcc phase rather than the other two possible phase transitions. A follow up  first-principles theoretical study challenged this conclusion and predicted that the hcp phase has lower enthalpy, than both the bcc and the fcc phases, in the 0-200 GPa pressure range and thus, should be  the high-pressure (HP) phase above 165 GPa \cite{Johansson1998}. However, the  $\alpha$-Mn phase was not included in the theoretical study and therefore the relative stability  of the hcp  over the $\alpha$-Mn phase remains an open question.

To date, the crystal structure of the HP phase of Mn  remains controversial: experimentally, the quality of the previous XRD study does not allow a definitive indexing of the HP phase and theoretically, the relative stability of the  $\alpha$-Mn and  hcp phases was not examined. In order to address these issues, we have carried out a  detailed x-ray powder diffraction  and computational study of $\alpha$-Mn up to 220 GPa. Surprisingly, no pressure induced phase transition was experimentally observed up to the highest pressure of this study and manganese remains in the $\alpha$-Mn phase. Interestingly, our DFT calculations predict that the hcp-Mn phase becomes lower in enthalpy than the $\alpha$-Mn phase above 160 GPa which suggests the hcp phase is energetically favorable at high pressure. The apparent discrepancy is attributed to  a significant kinetic energy barrier separating the $\alpha$-Mn phase from the hcp phase. Substantial electron transfer, as concluded from the result of  Bader charge analysis \cite{Bader1990}, between Mn atoms in the $\alpha$-Mn phase results in an ionic-like bonding and stabilizes this phase over the purely metallic hcp-Mn.

\section{Methods}

\subsection{Experimental methods}
High purity commercially available  (Sigma-Aldrich $>$99.9\%) fine powder of Mn was used for the angle dispersive XRD measurements.  The sample was loaded into a diamond-anvil cell (DAC) with neon (Ne) as a pressure transmitting medium (PTM). A Pilatus 1M CdTe detector was used at the undulator XRD beamline at GeoSoilEnviroCARS (sector13), APS, Chicago to collect pressure dependent X-ray diffraction data. The X-ray probing beam spot size was focused to approximately 2-4 $\mu$m, additional details on the XRD experimental setups are given in Ref. \onlinecite{Prakapenka2008}. Pressure was determined using a known room temperature  equation of state (EOS) of Ne \cite{Dewaele2008}.  Integration of powder diffraction patterns to yield scattering intensity versus 2$\theta$ diagrams and initial analysis were performed using the DIOPTAS program \cite{Prescher2015}. Calculated XRD patterns were produced using the POWDER CELL program \cite{Kraus1996}, for the corresponding crystal structures according to the EOSs determined experimentally and theoretically in this study and assuming continuous Debye rings of uniform intensity. Le Bail refinements were performed using the GSAS software \cite{Larson2000}. Indexing of XRD patterns has been performed using the DICVOL program \cite{Boutlif2004} as implemented in the FullProf Suite.

\subsection{Computational methods}
First-principles calculations were performed using the spin-polarized version of the Vienna \emph{ab initio} Simulation Package (VASP)\cite{Kresse1996}. Generalized gradient approximation (GGA) was employed with projected augmented wave (PAW) potentials,\cite{Kresse1999,Blochl1994} and Perdew-Burke-Ernzerhof (PBE) exchange correlation functional\cite{Perdew1996}. The wavefunctions were expanded in a planewave basis set with an energy cutoff of  300 eV. Valence electron configuration of 3p$^6$3d$^5$4s$^2$ for Mn atom was employed. Crystal structure search was carried out using the particle swarm-intelligence optimization (PSO) algorithm \cite{Wang2010,Wang2012}. All predicted crystal structures were processed through density functional theory calculations. Since we are interested in the high pressure (simple) phases of Mn, structure search was carried out at 100 GPa, 165 GPa, and 230 GPa with unit cells containing between 2 and up to 20 Mn atoms.  Lattice dynamical calculations were performed using the linear response Hessian matrix obtained using VASP program on a 7$\times$7$\times$7 q-point mesh and a 2$\pi$$\times$0.08 \AA$^{-1}$ k-point spacing and post processed using the PHONOPY code \cite{Togo2008}.

\section{Results and Discussion}
Figure 1 shows integrated diffraction patterns of Mn at selected pressures up to 220 GPa. No discontinuous changes (e.g. appearance of new Bragg peaks) are observed and all the main Bragg peaks of the $\alpha$-Mn phase are observed up to the highest pressure of this study. Moreover, no sign of the Bragg peak attributed to bcc-Mn by Fujihisa and Takemura (expected at 10.9$^o$ in the pattern of the inset of Fig. 1) is observed in any of the patterns in our study, and all observed Brag peaks can be indexed with the $\alpha$-Mn phase.

\begin{figure}[ht]
\begin{center}
{\includegraphics[width=\linewidth]{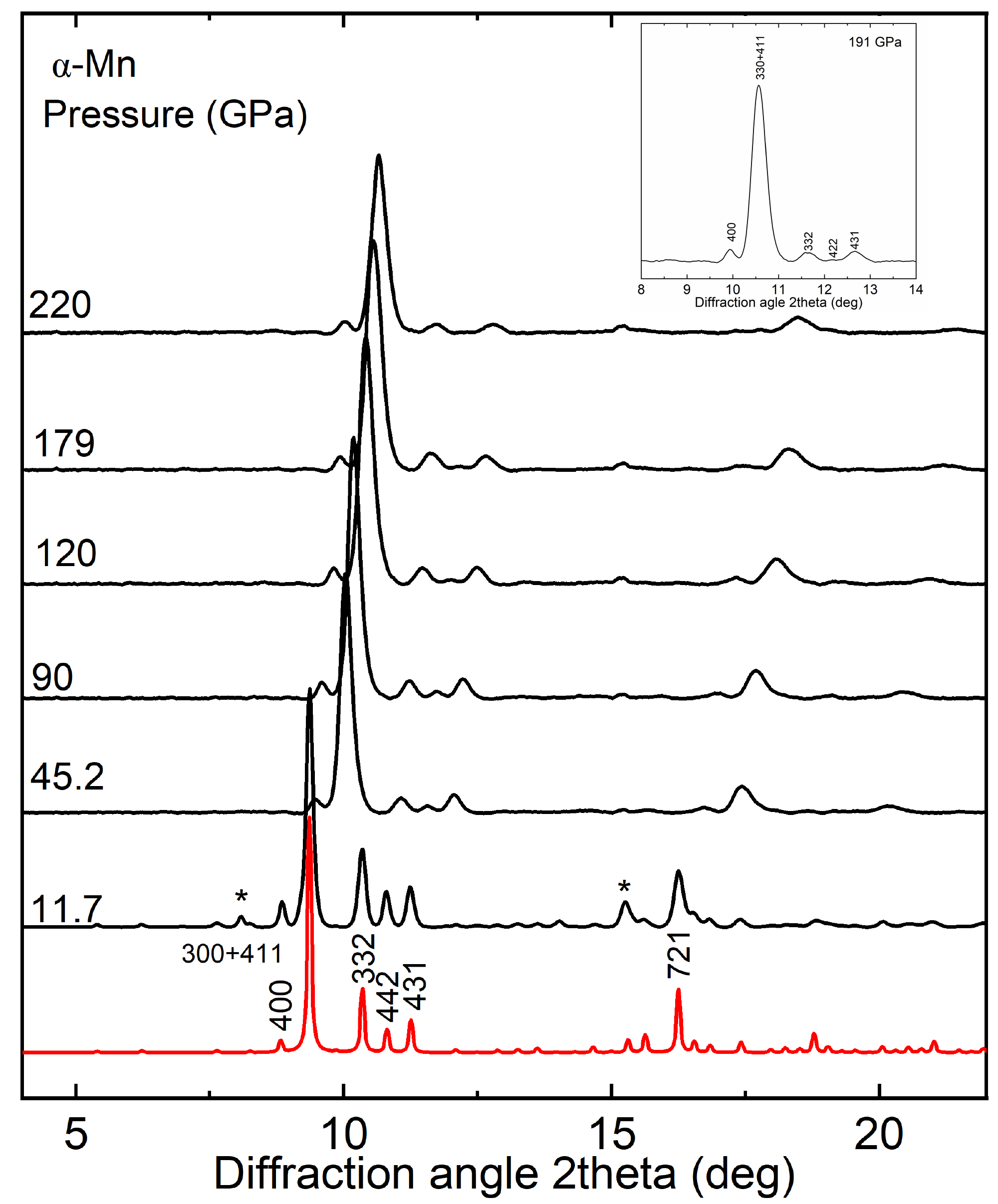}}
\caption{XRD patterns of $\alpha$-Mn at selected  pressures  on pressure increase. The red pattern is the calculated pattern of $\alpha$-Mn at 14 GPa. Miller indices of the main peaks are also noted. The asterisks mark  Bragg peaks from the rhenium gasket. The inset shows an expanded view of the pattern at 191 GPa. The X-ray wavelength is $\lambda$=0.3344\AA.}
\end{center}
\end{figure}

From the XRD data, the cell volume of the $\alpha$-Mn structure is determined as a function of pressure and compared with the values reported in  Ref.\onlinecite{Fujihisa1995}, see Fig. 2. Our obtained equation of state (EOS) implies a less compressible structure than the one of Ref. \onlinecite{Fujihisa1995}. We conducted unweighted fits to the experimental P-V data using a third-order Birch-Murnaghan EOS \cite{Birch1978a} and determined the bulk modulus $B$ and its first derivative $B'$  at zero pressure: $B_0$=204(3)GPa and $B_0'$=3.7(4). The determined in this study $B_0$ is substantially  higher while  $B_0'$ is lower than the ones determined by Fujihisa and Takemura using the Vinet EOS \cite{Vinet1987}: $B_0$=158(3)GPa and $B_0'$=4.6. Using a  fixed $B_0'$=4.6 and Vinet EOS, we determined a $B_0$=176(1) that is much closer to the value in Ref. \onlinecite{Fujihisa1995}. One possible explanation for the  difference in the determined elastic parameters might be the different methods used in the two studies for pressure determination; Ne EOS in our study \cite{Dewaele2008} and Fe (gasket) EOS in Ref. \onlinecite{Fujihisa1995}. Use of the gasket as pressure marker could provide an inaccurate pressure \cite{Anzellini2014,Dewaele2007}. We don't expect that the use of Ne as PTM in our study as opposed to a 4:1 mixture of methanol and ethanol in Ref. \onlinecite{Fujihisa1995} should play a role at these pressures \cite{Dewaele2006}.

\begin{figure}[ht]
\begin{center}
{\includegraphics[width=\linewidth]{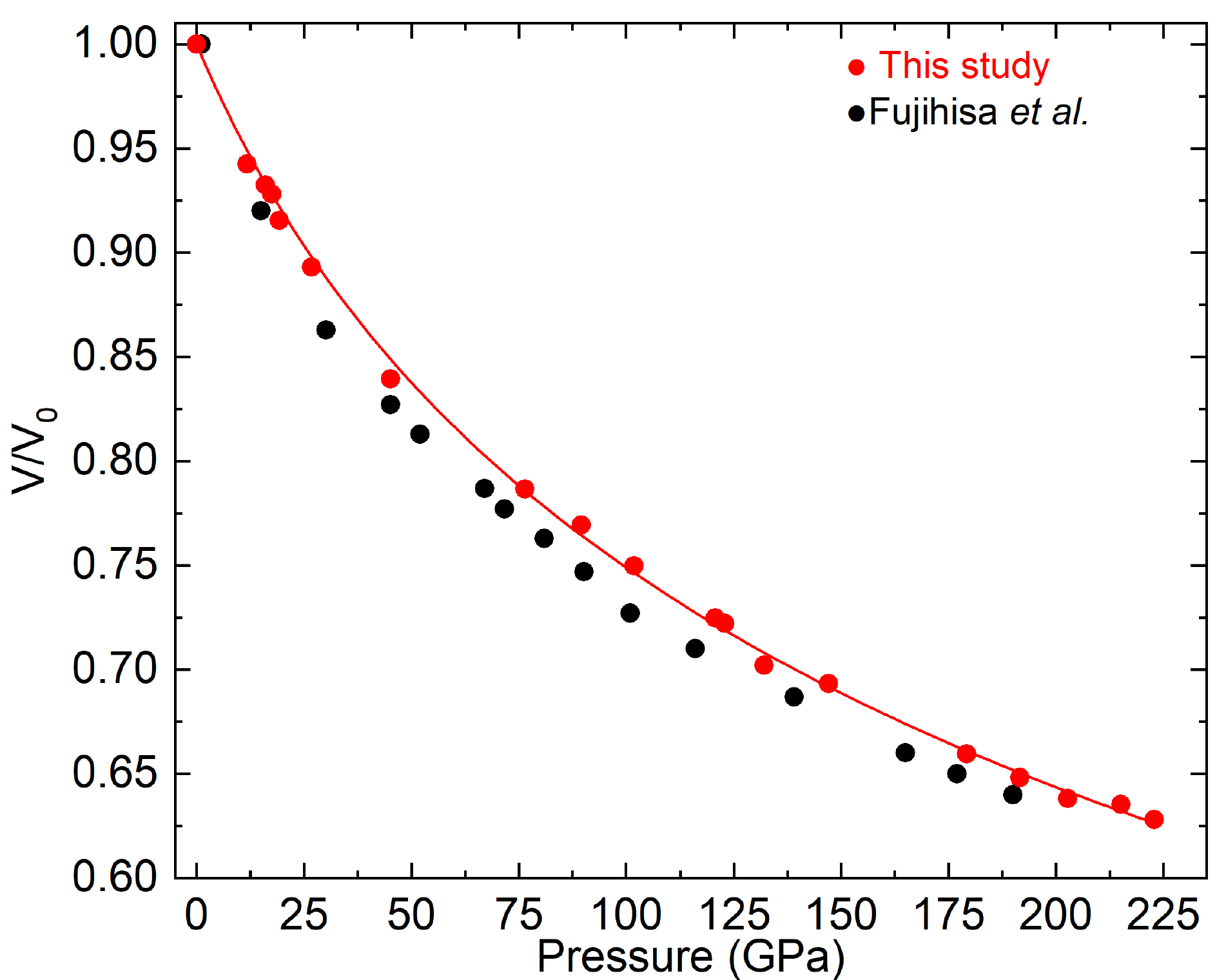}}
\caption{a)  Volume-pressure data for $\alpha$-Mn.  The solid curve is unweighted third-order Birch-Murnaghan EOS fit to the experimental data points \cite{Birch1978a}. The data  from  Ref. \onlinecite{Fujihisa1995} (black dots)  are also plotted for comparison.}
\end{center}
\end{figure}

Aiming to resolve the discrepancy concerning the presence of the additional Bragg peak in the study of  Fujihisa and Takemura, in Fig. 3(a) we compare the experimental XRD pattern of Mn at 190 GPa from Ref. \onlinecite{Fujihisa1995}  and the calculated pattern of hcp Fe at 190 GPa using the EOS of Ref. \onlinecite{Dewaele2006}. As it can be clearly seen, the new Bragg peak initially attributed to the 110 peak of bcc-Mn in Ref. 2, can be indexed with the 002 peak of hcp Fe (gasket) in addition to the other peaks already index with the 100 and 101 reflection of Fe. Moreover,  in the patterns of Ref.\onlinecite{Fujihisa1995}  the relative intensities between  the main peak of $\alpha$-Mn and the new peak remained the same with pressure increase. Based on these two observations it is plausible to conclude that the  Bragg peak observed above 160 GPa in Ref.\onlinecite{Fujihisa1995} is actually an additional (002) peak originated from the gasket and in reality, there is no pressure induced phase transition of Mn taking place, in agreement with our study.

\begin{figure}[ht]
\begin{center}
{\includegraphics[width=\linewidth]{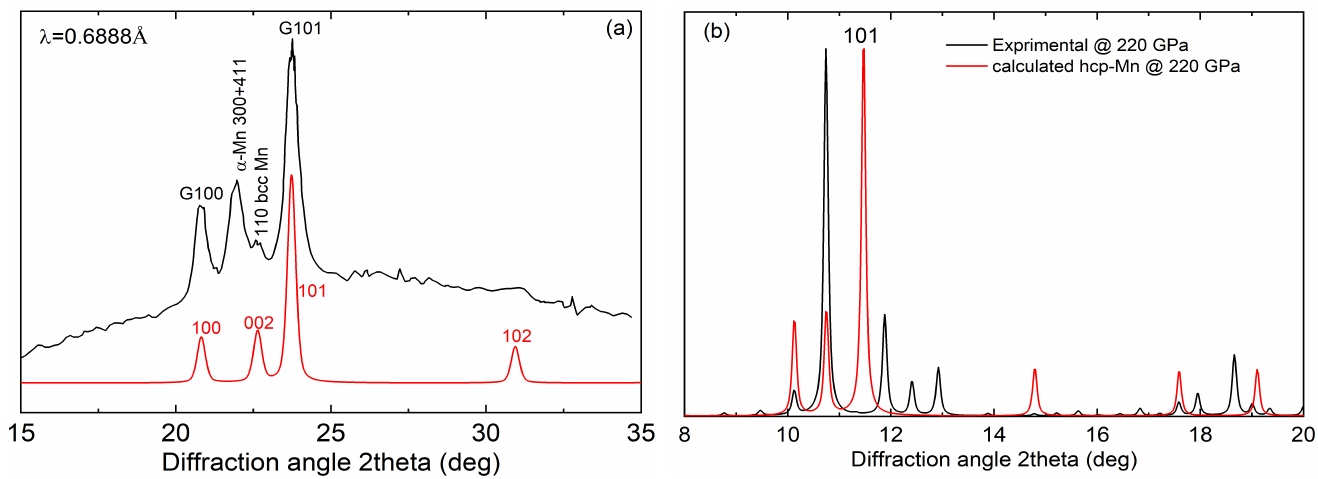}}
\caption{a) Comparison between the XRD pattern of Mn at 190 GPa from Ref. \onlinecite{Fujihisa1995} (black) and the calculated pattern of hcp Fe at 190 GPa (red) using the EOS of Ref. \onlinecite{Dewaele2006}. Miller indices of hcp Fe are denoted. Miller indices in the experimental pattern are initial indices taken from Ref.\onlinecite{Fujihisa1995}. The peaks indexed G100 and G101 are Bragg peaks from the Fe gasket. b) Comparison between the experimental XRD patten of this study (black) and the calculated pattern of hcp-Mn using the calculated EOS of this study (red).}
\end{center}
\end{figure}

The intuitive expectation on elemental metals under pressure is that they are all likely to favor close-packed structures (hcp, fcc and variants). At sufficient compression, when most valence electrons are pushed into interstitial sites, the crystal structure of elemental metals is mainly determined by the packing of the cations similar to the situation at ambient conditions and therefore close-packed structure would emerge. To investigate this possibility, we carried out crystal structure search at 100 GPa, 165 GPa and 230 GPa that bracket the transition pressure reported by Fujihisa and Takemura \cite{Fujihisa1995} and the maximum pressure reached in this study. The hcp-Mn (SG: $P6_3/mmc$) was found to be the most energetically favorable phase followed by fcc-Mn (SG: $Fm-3m$) and the bcc-Mn (SG: $Im-3m$) was the least stable. This finding is consistent with the results of Johansson et al. \cite{Johansson1998} where the hcp phase was reported as the most stable one compared with the bcc and fcc structures. Specifically, the calculated relative enthalpies of the $\alpha$-Mn and hcp-Mn as a function of pressure (Fig. 4) show that the hcp phase becomes lower in enthalpy than the α-Mn above 165 GPa, implying a possible phase transition above this pressure. For this reason in Fig. 3(b) we compare the experimental pattern of this study with the calculated pattern of hcp-Mn using the EOS of hcp-Mn calculated in this study. Obviously, there is a strong discrepancy between the two patterns thus, ruling out the presence of hcp-Mn, even at the level of coexistence with α-Mn (the strongest 101 peak of hcp-Mn cannot be observed).

\begin{figure}[ht]
\begin{center}
{\includegraphics[width=\linewidth]{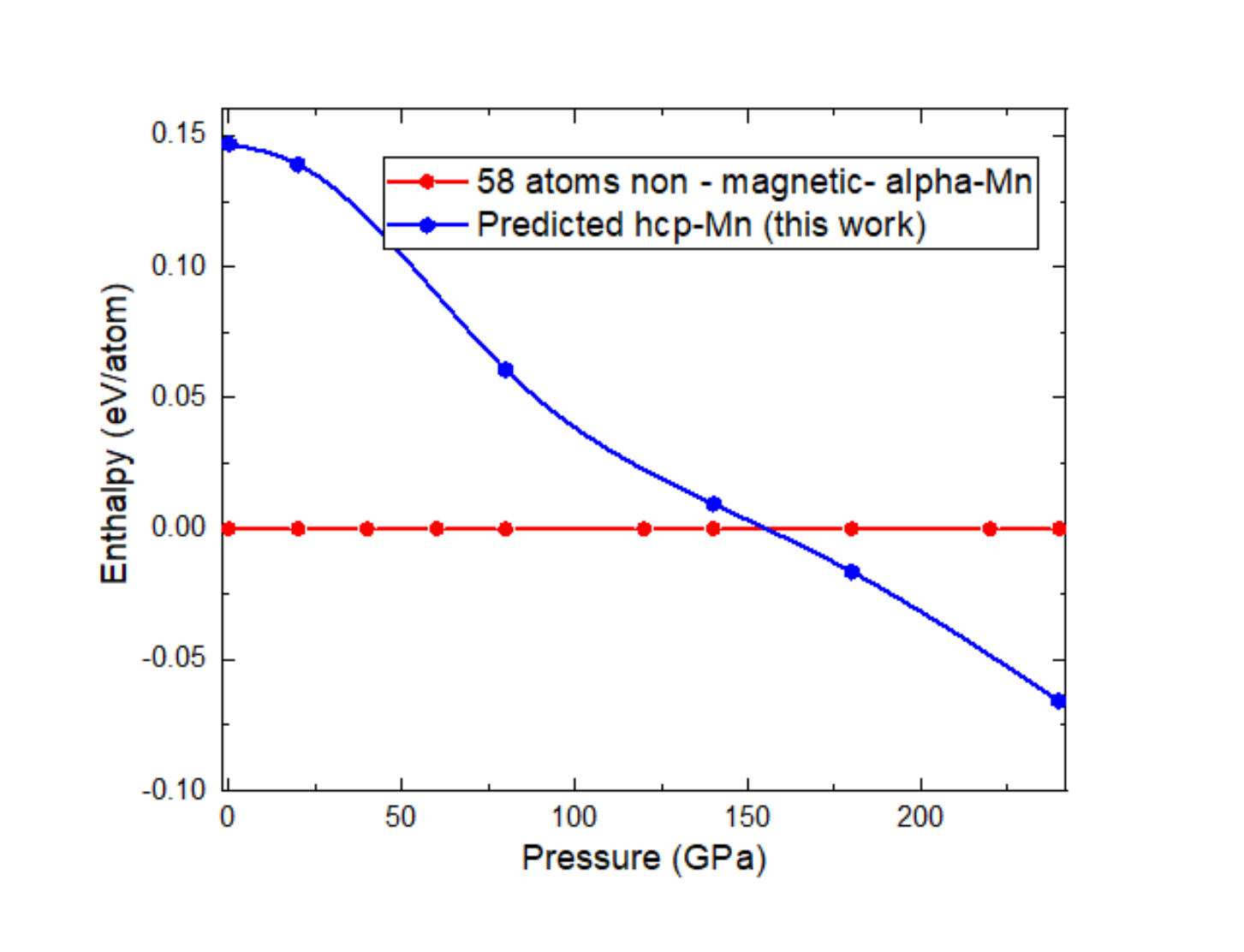}}
\caption{Calculated enthalpy differences for $\alpha$-Mn and hcp-Mn  phases as a function of pressure. The enthalpy of the $\alpha$-Mn phase is taken as the reference. }
\end{center}
\end{figure}

Below the melting point of a material, the lattice contribution to the free energy ($F_{vib}$) can play an important role in the relative stability of two different phases \cite{Pavone1998,Adeleke2017}. We therefore analyze the effects of lattice vibration on both $\alpha$-Mn and hcp-Mn at a pressure (180 GPa) that is slightly above the calculated $\alpha$-Mn $\rightarrow$ hcp-Mn transition. The calculated $H+F_{vib}$ for the $\alpha$-Mn and hcp-Mn at 180 GPa are shown in Fig. 5(a) as functions of temperature. The enthalpies ($H$) are taken from the static calculation (Fig. 4). The hcp-Mn is generally more stable than $\alpha$-Mn up to the maximum temperature (1500 K), indicating the absence of any lattice dynamic or thermally induced phase transition.

The calculated temperature dependence of the vibrational entropies ($S$) of the two phases are shown in Fig. 5(b). Below 160 K, the entropy of the $\alpha$-Mn is slightly higher than the hcp-Mn. Although, in this temperature range, the entropic contribution to the free energy difference is finite and in favor of the $\alpha$-Mn (see Fig. 5(b) inset), it is much less than the internal energy difference and therefore cannot induce a phase transition from hcp-Mn $\rightarrow$ $\alpha$-Mn. In fact, at 0 K, hcp-Mn is 0.020 eV/atom more stable than $\alpha$-Mn. In the same vein, at sufficiently high temperature (above 160 K) below the melting temperature of Mn, hcp-Mn has slightly higher entropy and is progressively more stable than the $\alpha$-Mn, e.g., 0.023 eV/atom more stable than α-Mn at 1500 K. Thus, in the pressure range of interest, the lack of transition to hcp-Mn is not likely due to temperature effects.

\begin{figure}[ht]
\begin{center}
{\includegraphics[width=\linewidth]{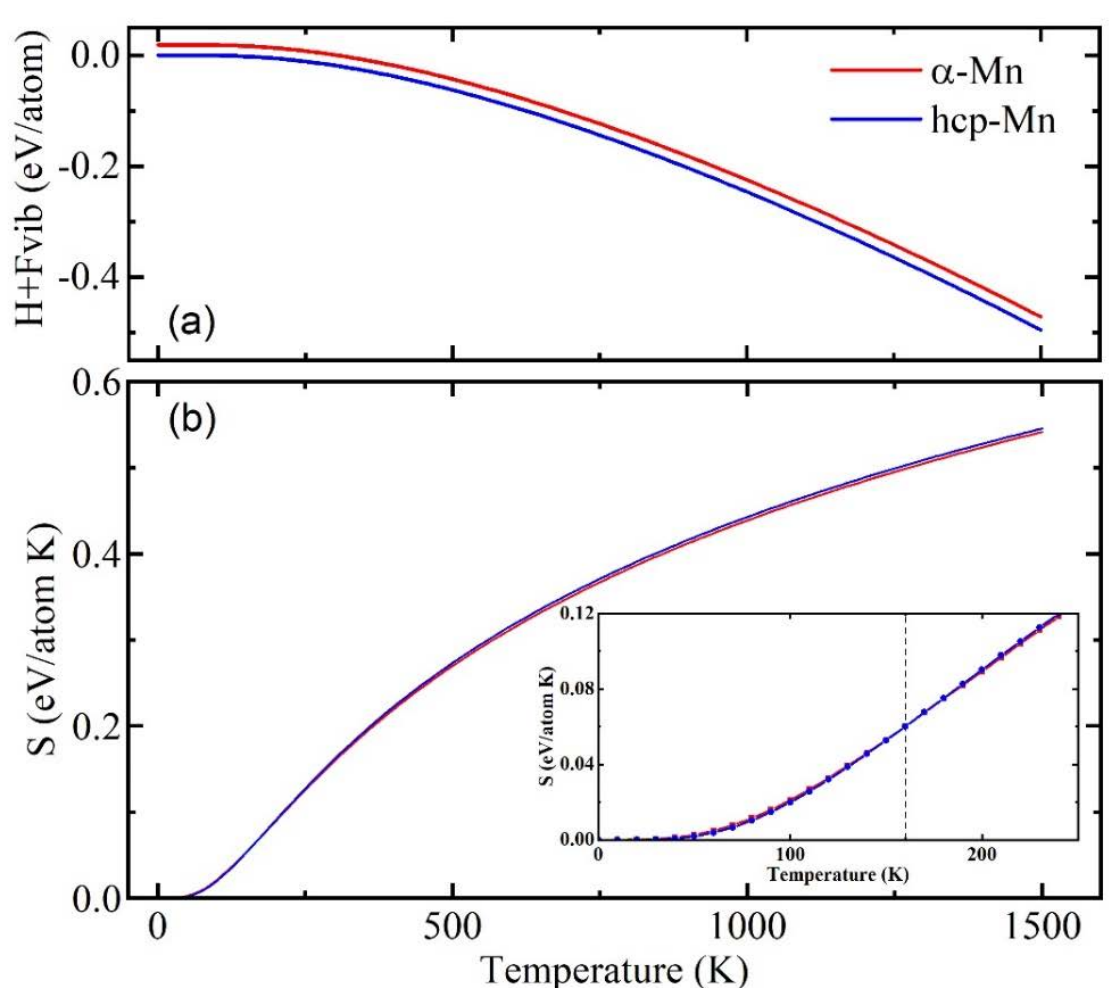}}
\caption{The temperature dependent (a) $H+F_{vib}$ for the $\alpha$-Mn and hcp-Mn at 180 GPa (see text). The enthalpy of the hcp-Mn at 0 K was used as the zero-point reference. (b) Vibrational entropies of the $\alpha$-Mn and hcp-Mn at 180 GPa. The inset shows the hcp-Mn having slightly lower vibrational entropies below 160 K.}
\end{center}
\end{figure}

To gain further insight into the extended pressure stability of the $\alpha$-Mn phase and more importantly to understand the discrepancy between experimental and theoretical results we examined alternative mechanisms that can stabilize the $\alpha$-Mn phase. In this context,  we have performed  charge transfer calculations in both the  $\alpha$-Mn and hcp-Mn phases as a function of pressure using Bader analysis \cite{Bader1990}. In the case of the hcp-Mn, no charge transfer between Mn ions was observed for the whole pressure range of this study. On the other hand, a surprisingly substantial charge transfer is predicted for Mn ions in the $\alpha$-Mn phase above 100 GPa for all the four crystallographically inequivalent  Wyckoff positions (2a, 8c, 24g and 24g). As it can be clearly seen in Fig. 6, the charge transfer is low up to 100 GPa, followed by an abrupt first-order-like increase above this pressure.

\begin{figure}[ht]
\begin{center}
{\includegraphics[width=\linewidth]{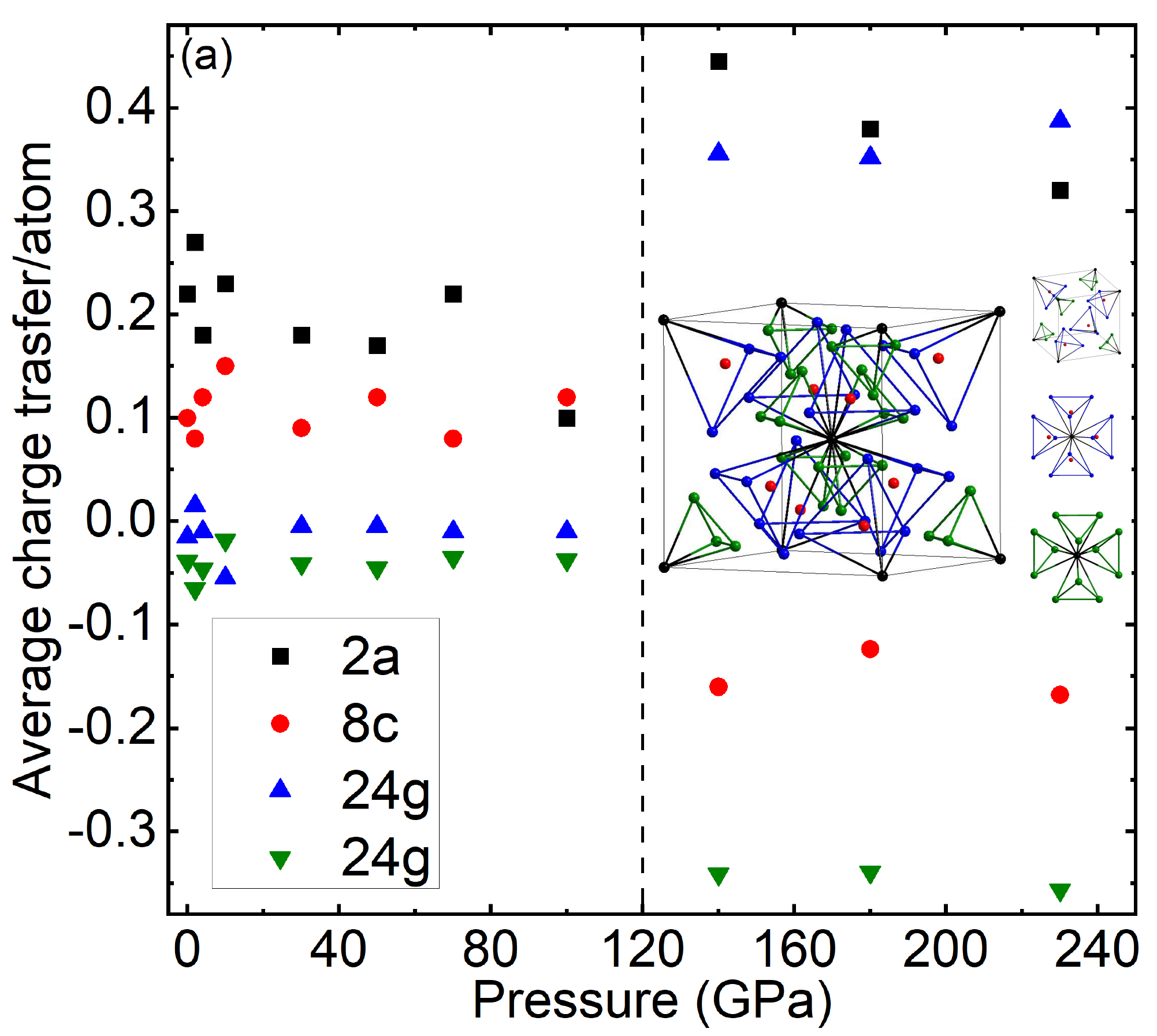}}
\caption{ (a)average per atom   and (b) sum ($i.e.$ average per atom multiplied by the site multiplicity) Bader charge transfer of Mn ions in the corresponding Wyckoff positions as a function of pressure as calculated in this study .}
\end{center}
\end{figure}

This effect can be described as a pressure-induced autoionization of the Mn atoms, forming a self-salt, in the $\alpha$-Mn phase. The $\alpha$-Mn structure is arranged in a cubic unit cell with 2a atoms occupying the corners and center of the cube. The first set of 24g atoms are split into eight 3-atom groups, which form 4 tetrahedrons with 4 cornered 2a atoms and a cluster of 4 tetrahedrons sharing one vertex made of the centered 2a atom. The 8c atoms occupy the voids of 8 tetrahedrons, forming nominal 8 [Mn$_5$] entities. These filled tetrahedrons have large Mn-Mn contacts, ranging between 3.5-3.9\AA (ambient value). The second set of 24g atoms also split into eight 3-atom groups, and form empty tetrahedrons with the 2a atoms exactly the same way. The empty tetrahedrons have smaller size (Mn-Mn distance: 2.1-2.4 \AA), nominally [Mn$_4$] entities. Another way of explaining the structure is that in the bcc unit cell the 8 cornered Mn atoms each forms a tetrahedron with 3 Mn atoms, and the centered Mn atom forms a cluster of 8 tetrahedrons with 24 Mn atoms sharing the center vertex. The unit cell therefore contains 16 tetrahedrons but only 8 are filled by 8c atoms, resulting in alternating small (empty) and large (filled) 3D extension of tetrahedrons, see inset in Fig. 6. The calculated Bader charges (Fig. 6) reveal substantial charge transfer from the small tetrahedrons to large tetrahedrons nearby, forming self-ionized [Mn$_4$]$^{\delta+}$ [Mn$_5$]$^{\delta-}$ pairs. Interestingly, the abrupt increase of the  charge transfer happens slightly below the pressure (165 GPa) that the hcp-Mn becomes lower in enthalpy than the $\alpha$-Mn phase.    Thus, it is plausible to conclude that the ionic (in addition to the metallic) character of the bonding in the $\alpha$-Mn phase above 120 GPa, as opposed  to the purely metallic  bonding in the hcp-Mn phase, is the driving reason for the stability of the $\alpha$-Mn phase.

The occurrence of such strong charge transfer in an elemental solid is rare, but not unprecedented. Early theoretical study suggests that solid hydrogen can polarize into H$^+$H$^‒$ molecules under pressure \cite{Edwards1997}. Subsequently, self-ionization was observed in several elemental solids at high pressure, such as in the case of the insulating $hP4$ phase of Na \cite{Ma2009}, $\gamma$-phase of B \cite{Oganov2009}, or the predicted N$_6$ allotrope \cite{Greschner2016} and the all-nitrogen metallic salt \cite{Sun2013}. In these cases ionic bonding involves either localized electrons (electride phases of alkali metals) or molecular-like (e.g. N$_2^{\delta+}$ and N$_5^{\delta-}$) entities. In general, ionicity in elemental solids is the result off many-body interactions, which are stronger in squeezed volumes when orbital splitting is enhanced. The $\alpha$-Mn structure is made of two tetrahedrons with very different electronic structures. Under high pressure, the filled [Mn$_5$] tetrahedron becomes anionic and more stable, because its neutral state would have unoccupied bonding orbital. This orbital creates an acceptor band below the Fermi level that receives the electrons transferred from the [Mn$_4$] tetrahedron nearby. After this, substantial nearest-neighbor electrostatic attraction arises between [Mn$_4$]$^{\delta+}$ [Mn$_5$]$^{\delta-}$, resulting in a very high cohesive energy making the $\alpha$-Mn phase metastable in an extended pressure range.

Another interesting point to note is that from the crystallographic point of view, the $\alpha$-Mn and hcp-Mn phases are not linked by either group-subgroup relation or a common subgroup. This means that the $\alpha$-Mn $\rightarrow$ hcp-Mn phase transition, if ever happens, will likely proceed in an abrupt manner with no clearly defined order parameters \cite{Stokes2002}. Such a transition is associated with a very high activation barrier compared to a displacive reconstructive phase transition. However, since the phase transition path cannot be uniquely described (by collective variables), we cannot estimate the energy barrier for $\alpha$-Mn $\rightarrow$ hcp-Mn transition. From the above discussion it is plausible to conclude that the extended stability of the $\alpha$-Mn could be attributed to both the ionic character of this phase above 120 GPa and also the presumably high energy barrier. This kind of dynamic stability is commonplace for materials under high pressure, where interesting crystal structures/stoichiometries were found stable beyond their pressure range of stability \cite{Adeniyi2020}.

Our study calls for follow-up experimental, e.g.  XANES might provide experimental evidence of the charge transfer although the critical pressure  is relatively high for such measurements \cite{Levy2020},   and theoretical studies aiming to  fully elucidate the mechanism of the stability of the $\alpha$-Mn phase. Our Bader analysis reveals that under pressure, in addition to electride phases and molecular-like salts,   ionic bonding in  elementals solids could also  exists between exclusively  monatomic ions of the same metallic element.

\begin{acknowledgments}
This work was performed under the auspices of the U. S. Department of Energy by Lawrence Livermore National Security, LLC under Contract DE-AC52-07NA27344. This work was supported by the DARPA (Grants No. W31P4Q1310005 and No. W31P4Q1210008), the Deep Carbon Observatory DCO, and Natural Sciences and Engineering Research Council of Canada (NSERC). This research used resources of the Advanced Photon Source, a U.S. Department of Energy (DOE) Office of Science User Facility operated for the DOE Office of Science by Argonne National Laboratory under Contract No. DE-AC02–06CH11357. GSECARS is supported by the U.S. NSF (EAR-1128799) and DOE Geosciences (DE-FG02-94ER14466). Use of the COMPRES-GSECARS gas loading system was supported by COMPRES under NSF Cooperative Agreement EAR -1606856 and by GSECARS through NSF grant EAR-1634415 and DOE grant DE-FG02-94ER14466.  The authors thank Sergey N. Tkachev for help gas loading samples at the APS, Sector-13 GSECARS beamline. E.S. thanks K. Syassen for fruitful discussions and for a critical reading of the manuscript. E.S. was partially supported by  LDRD grant 18-LW-036.
\end{acknowledgments}

merlin.mbs apsrev4-1.bst 2010-07-25 4.21a (PWD, AO, DPC) hacked

\end{document}